\documentclass[prd,preprintnumbers,12pt]{revtex4}
\pdfoutput=1
\usepackage{epsfig}
\usepackage{graphicx}
\usepackage{bm}

\newcommand{\be}{\begin{equation}}
\newcommand{\dd}{\displaystyle}
\newcommand{\ee}{\end{equation}}
\newcommand{\bea}{\begin{eqnarray}}
\newcommand{\eea}{\end{eqnarray}}

\newcommand{\nn}{\nonumber}
\newcommand{\de}{\partial}
 
\def\nn{\nonumber}
\def\de{\partial}

 \def\slash#1{\setbox0=\hbox{$#1$}#1\hskip-\wd0\dimen0=5pt\advance
       \dimen0 by-\ht0\advance\dimen0 by\dp0\lower0.5\dimen0\hbox
         to\wd0{\hss\sl/\/\hss}}
\def\be{\begin{equation}}
\def\dd{\displaystyle}
\def\ee{\end{equation}}

\def\bea{\begin{eqnarray}}
\def\eea{\end{eqnarray}}
\def\tx{\tilde x}
\def\tde{\tilde\de}
\def\7{\tilde}
\def\8{\hat}

 \def\slash#1{\setbox0=\hbox{$#1$}#1\hskip-\wd0\dimen0=5pt\advance
       \dimen0 by-\ht0\advance\dimen0 by\dp0\lower0.5\dimen0\hbox
         to\wd0{\hss\sl/\/\hss}}

\def\tx{\tilde x}

\usepackage{color}

\begin{document}


\title{Confined dynamical systems with Carroll and Galilei symmetries}
\author{Andrea Barducci}\email{barducci@fi.infn.it} \affiliation{Department of Physics, University of Florence and INFN Via G. Sansone 1, 50019 Sesto Fiorentino (FI), Italy}
\author{Roberto Casalbuoni}\email{casalbuoni@fi.infn.it}
\affiliation{Department of Physics, University of Florence and INFN Via G. Sansone 1, 50019 Sesto Fiorentino (FI), Italy}
\author{Joaquim Gomis}\email{gomis@ecm.ub.es}\affiliation{Departament d'Estructura i Costituents de la Mat\`eria}\affiliation{Departament de F\'isica,
Universitat de Barcelona, Diagonal 647, 08028 Barcelona, Spain and
C.E.R. for Astrophysics, Particle Physics and Cosmology, Barcelona,
Spain}
\date{\today}

\begin{abstract}
We introduce a general method to construct,  directly in configuration space, classes of dynamical systems invariant under generalizations of the Carroll and of the Galilei groups.
The method does not make use of any non-relativistiv limiting procedure, although the starting point is a lagrangian Poincar\'e invariant in the full space. It consists in considering  a space-time in $D+1$ dimensions and partitioning it in two parts, the first Minkowskian and the second Euclidean. The action consists of two terms
that are separately invariant under the Minkowskian and Euclidean partitioning. One of those contains a system of lagrangian multipliers that confine the system to a subspace. The other term defines the dynamics of the system. The total lagrangian is invariant under the  Carroll or  the Galilei groups with zero central charge.

\end{abstract}


\maketitle


\section{Introduction}

During the last few years there has been interest in the "non-relativistic"  Carroll 
\cite{Levy-Leblond} and Galilei groups.
 One of the reasons for this interest is the relevance of BMS symmetry \cite{BMS-1}, which is related to the conformal Carroll group \cite{Duval:2014}, to flat space holography
 \cite{Banks:2003vp,deBoer:2003vf,Arcioni:2003xx,Barnich:2010eb,Bagchi:2012cy}
 . Another reason is  the non-relativistic holography as a tool to study strongly interacting field theories in condensed matter systems \cite{review,review1}. The associated non-relativistic gravities
like Newton-Cartan \cite{Cartan:1923zea},
  Horava \cite{Horava:2009uw}, stringy Newton Cartan  \cite{Bagchi:2009my,Andringa:2012uz},
torsional Newton Cartan \cite{Christensen:2013lma}
or Schr\"odinger \cite{Bergshoeff:2014uea}, Carroll 
\cite{Henneaux:1979vn,Hartong:2015xda,Bergshoeff:2017btm} have been studied.
There is  matter coupled to non-relativistic gravity, Carroll gravity, for example, for particles  \cite{Kuchar:1980tw,Kluson:2017pzr,Barducci:2017mse,Bergshoeff:2014jla,Duval:2014uoa},  and for  extended objects \cite{Andringa:2012uz,Kluson:2017pzr}  for
Galilean and for Carroll field theories \cite{Jensen:2014aia,Hartong:2014pma} coupled to a Newton-Cartan and Carroll \cite{Bergshoeff:2017btm}
background. 

 Since   most  of the models invariant under Carroll or Galilei group present in the literature are described by an action  in phase space, the main purpose of this paper is to propose a general method to construct the corresponding action  in
configuration space. As we shall see, this description corresponds to a system confined in a particular region of the space-time. 
 The corresponding lagrangian systems will be {\it invariant} under generalizations of the Carroll and of the Galilei groups, 
\cite{Gomis:2000bd,Gomis:2005pg,Brugues:2004an,Brugues:2006yd,Bergshoeff:2014jla,Batlle:2016iel,Cardona:2016ytk}.

The Galilei and Carroll  groups 
can be obtained via a convenient contraction of the Poincar\'e group \cite{Bacry:1968zf}. In order to obtain the Bargmann group \cite{Bargmann:1954gh}, that is the Galilei group with a central charge, 
it is necessary to extend the Poincar\'e group by a $U(1)$ factor, whereas  in the case of three dimensions one needs to consider the contraction of Poincar\'e $\otimes U(1)\otimes  U(1)$
\cite{Bergshoeff:2016lwr}, 
since in this case the Galilei group has two central extensions \cite{LL}. Otherwise one gets the Galilei algebra with zero central charge. 
It should be noted that the "non-relativistic" limit is not unique \cite{Gomis:2000bd,Danielsson:2000gi,Gomis:2004pw,Batlle:2016iel}
when we have in mind extended objects. In other words, there is not a unique contraction of the Poincar\'e  group. In some cases these new algebras could contain non-central extensions \cite{Brugues:2004an,Brugues:2006yd}. 
 From now on, unless differently specified, for Galilei and Carroll algebras, we will mean the algebras with zero central or non-central charges.

The general structure of the contracted algebras is that both contain the direct product of a Poincar\'e algebra in lower dimensions times an Euclidean algebra of rotations and translations in the complementary 
 spatial
dimensions. The contracted algebras contain also generalized boosts that rotate
the generators of the lower Poincare  with the ones of the Euclidean group.
These boosts are  a generalization of the ones of
Galilei  and  Carroll groups.
We will write  the generators of the  the  Poincar\'e group, $ISO(1,D)$, $M_{\mu\nu}$ and $P_\mu$ $\mu,\nu=0,1,\cdots,D$,  as $(M_{\alpha\beta},M_{ab},P_\alpha, P_a, M_{\alpha b}\equiv B_{\alpha b})$, with $\alpha,\beta=0,1,\cdots,k-1$ and $a,b=k,\cdots,D$.

 The contractions will be defined  dividing the components of the momentum $P_\mu$,  in two sets $P_\alpha$ and $P_a$. The  contractions we will consider consist,  in the Carroll case, in rescaling  the momenta $P_\alpha$ by a factor $1/\omega$ and then taking the limit $\omega\to\infty$, 
 explicitly $ \tilde P_\alpha= P_\alpha/\omega$, where  $\tilde P_\alpha$ will be the
Carrollian generators.
 At the same time, we will need to rescale the boosts by the same factor. The same procedure is followed in the Galilei case, but this time we rescale the $P_a$'s as
 $\tilde P_a=P_a/\omega$.
The boosts  work differently in the two cases.  More precisely, in the Carroll case the commutator of the boosts with
 $P_a$ is proportional to $P_\alpha$, whereas in the Galilei case the two momenta are exchanged. This is the distinctive feature between the two contracted algebras.

An analogous procedure in space-time 
 can be
followed 
by considering a Minkowski space-time $M(1,D)$, and partitioning it in two pieces, as the direct sum    $M(1,k-1)\oplus E(D+1-k)$, corresponding to the partition of the momentum
generators considered above.
In this case we will consider the realization of the Poincar\'e generators in terms of vector fields operating on the space-time,  $M(1,D)$, variables. Then, we will define two types of contractions by rescaling the  relativistic coordinates $x^\alpha$ by a factor ${\omega}$, that is $\tilde x^\alpha=\omega x^\alpha$,
in the Carroll case  and,  by the same  factor $\omega$,  the coordinates $x^a$,
 $\tilde x^a=\omega x^a$,
 in the Galilei case. The boosts connect the coordinates $x^\alpha$ to the $x^a$'s for the Carroll contraction  and the contrary happens for the Galilei one.

The previous  procedure can be repeated, by varying $k$,  $D$ times, not $D+1$, since rescaling all the momenta is equivalent to no rescaling. Notice that  the commutation relations among the momenta and the generators of the Lorentz group are linear in the momenta, and therefore the commutation relations do not change for an overall rescaling of the momenta. Therefore, we may have $D$ contractions of Carroll type and $D$ contractions of Galilei type, in total $2D$ possible contractions. We will define as a $k$ contraction of Carroll type, the contraction described above, and by the same rule a contraction of Galilei type. Notice that in the previous literature these contractions where called p-brane contractions. 

A $k$ contraction of Carroll type and a $D+1-k$ contraction of  Galilei type are dual one to the other. In fact, they can be obtained  one from  the other by exchanging the role of the 
 boosts.
In the case of $k=1$, this  duality is close to the one considered  in \cite{Duval:2014uoa} although not quite the same.

Models invariant under the Carroll or the Galilei group have been previously obtained by taking convenient "non-relativistic" limits on a relativistic action describing the original system in the D+1 dimensional Minkowski space-time together with  an appropriate rescaling of the parameters appearing in the action
\cite{Gomis:2000bd,Danielsson:2000gi,Gomis:2004pw,Gomis:2005pg,Brugues:2004an,Brugues:2006yd,Bergshoeff:2014jla,Batlle:2016iel,Cardona:2016ytk}.
Some of these models without central extensions can be recovered by our approach.

Our general procedure   in configuration space does not need to take
any "non-relativistic" limit
 to construct an invariant model under Carroll or Galilei.
  Starting with the Carroll case, we will  first consider an action invariant under   $ISO(1,D)$,   
  then we confine the system to a $M(1,k-1)$ subspace.
 The resulting action is automatically invariant under $ISO(1,k-1)$ and  depends on the coordinates $x^\alpha$ only. 
 
 The model is trivially invariant under the Euclidean sub-algebra and  can be made invariant under the boosts adding convenient terms in the euclidean sector of the total space-time. Since the Carroll boosts map the Minkowski coordinates into the Euclidean ones,  the corresponding variation of the action is linear in the euclidean coordinates and in their derivatives.
Therefore we can construct compensating terms using   lagrange multipliers times the coordinates of the Euclidean space
and their derivatives such to conserve the momentum also in that sector\label{note:1}. We will be more specific in the following. In fact,  the action might  depend explicitly on the coordinates, but it needs to conserve the momentum. This allows to construct linear combinations of the euclidean coordinates, 
We shall see that the lagrangian multipliers of the time derivatives of the euclidean coordinates can be thought as the
canonical momenta of  the Euclidean space variables.
The variation induced by the boosts in the action describing the dynamical  system is then compensated by a corresponding variation of the lagrange multipliers. On the other hand, the equations of motion resulting from the variation of the lagrange multipliers confine the system to live in the 
 Minkowski subspace,  since the typical implication is that the time derivatives of the euclidean variables $x^a$ vanish confining  the system  in the Minkowski subspace and to not propagate in the euclidean part. For this reason we will call the part of the action relative to the euclidean subspace as $S_{\rm confining}$.  Notice that in this procedure we  do not need to scale the  parameters appearing in our lagrangian.
We will discuss in a detailed way the construction  of  discrete 
 and  continuous dynamical systems following what outlined before.

The same considerations can be made for dynamical systems invariant under the Galilei algebra. In fact, this time we will introduce an invariant action under the Euclidean group, $E(D+1-k)$, depending on the variables $x^a$ only. In this case the Galilei boosts map the variables $x^a$ into the $x^\alpha$'s. and we can get a model invariant under the full Galilei group, by the addition of a confining action consisting of lagrange multipliers times convenient  combinations of the $x^\alpha$'s.  Note that the mass-shell
constraints of our models depend on momenta and coordinates on the longitudinal
 or on the transverse variables, but not depending on both.
Several of the previous constructed models do depend on all the variables, see for
example \cite{Gomis:2004ht,Batlle:2016iel,Cardona:2016ytk},
and therefore
can not be obtained from our procedure.

Strictly speaking, the Galilei invariant  models obtained by this procedure are not  dynamical 
in space-time, since the action of these systems does not  contain the time variable. This means that the dynamics happens only at a given time. In this sense, these models describe a sort of generalized instantons. The evolution in euclidean space can be obtained
in terms of an euclidean coordinate and the associated hamiltonian.
In general, it  is  possible to obtain interesting results. For instance, considering the dual of the 1-contraction of Carroll type (the Carroll particle), that is a $D$-contraction of Galilei type, it is possible to obtain, in configuration space, a model equivalent to the one, obtained  in \cite{Batlle:2017cfa} as the limit of a relativistic tachyon, in the phase space, for $c\to\infty$.  This model 
 has been called a Galilean massless particle \cite{Souriau}, referring to the invariance under the Galilei algebra with zero central charge. This model describes an euclidean non-relativistic instanton.


The organization of the paper is as follows: in Section \ref{sec:2} we discuss the $k$ contractions at the level of the algebra, whereas in Section \ref{sec:3} the same analysis is performed in configuration space. In Section \ref{sec:4} we present, in general, the construction of dynamical models invariant under Carroll and Galilei, in the discrete and in the continuous case,
as, for instance,  extended objects. In Section \ref{sec:5} we will discuss various explicit examples of  our procedure,  
ranging from the Carroll particle to massless particles up to a Carroll string
In the last Section  we present our conclusions and discuss possible extensions of the present work.

\section{$k$-contractions of the Poincar\'e group}\label{sec:2}

 Let us start considering the algebra of the Poincar\'e group in $D+1$ dimensions,   $ISO(1,D)$

 \bea
 \left[M_{\mu\nu},M_{\rho\sigma}\right]&=&-i(\eta_{\mu\rho}M_{\nu\sigma}+\eta_{\nu\sigma}M_{\mu\rho}
 -\eta_{\mu\sigma}M_{\nu\rho}
  -\eta_{\nu\rho}M_{\mu\sigma}),
\nonumber \\
  \left[M_{\mu\nu},P_{\rho}\right]&=&-i(\eta_{\mu\rho}P_{\nu}-\eta_{\nu\rho}P_{\mu}),
\nn\\
  \left[P_{\mu},P_{\nu}\right]&=& 0 \ ,\label{eq:1.1}
\eea
where $\mu=0,1,...,D$, $\eta_{\mu\nu}=(-;+,\cdots,+)$. 

Then, consider  the following two subgroups of $ISO(1,D)$: the Poincar\'e subgroup in $k$ dimensions, $ISO(1,k-1)$ and the euclidean group of roto-translations in $D+1-k$ dimensions, generated respectively by
\be
ISO(1,k-1):~~~M_{\alpha\beta},~~P_\alpha,~~~\alpha,\beta =0,1,\cdots,k-1,
\ee
\be
ISO(D+1-k):~~~M_{ab},~~P_a,~~~a,b =k,\cdots,D.
\ee
In these notations the generators of $ISO(1,D)$ are
\be
ISO(1,D):~~~M_{\alpha\beta},~~~M_{ab},~~P_\alpha,~~~P_a,~~~M_{\alpha b}\equiv B_{\alpha b}.
\ee
 The generators of $ISO(1,k-1)$ satisfy the algebra (\ref{eq:1.1}), with the indices $\mu.\nu,...$ replaced by $\alpha,\beta,\cdots$. Also the generators of $ISO(D+1-k)$ satisfy  the algebra (\ref{eq:1.1}), with the indices $\mu.\nu,...$ replaced by $a,b,,\cdots$ and with the replacement of the Minkowski metric with the euclidean one, $\eta_{ab}=\delta_{ab}$. Furthermore, the two subalgebras commute. Now let us study the behaviour of the boosts under the two subalgebras.
 We have
 \be
 ISO(1,k-1): [M_{\alpha\beta}, B_{\gamma c}]= -i(\eta_{\alpha\gamma}B_{\beta c}-\eta_{\beta\gamma} B_{\alpha c}),~~~[B_{\alpha a},P_\beta]=-i\eta_{\alpha\beta}P_a,\label{eq:2.5}
 \ee
 \be
 ISO(D+1-k): [M_{ab}, B_{\gamma c}]= -i(\eta_{ac} B_{\gamma b}-\eta_{bc}B_{\gamma a}),~~~[B_{\alpha a},P_b]=i\eta_{ab}P_\alpha.\label{eq:2.6} \ee
 We see that the boosts behave like vectors under both groups. Finally, the commutator among boosts is given by
 \be
[B_{\alpha a},B_{\beta b}]=-i(\eta_{\alpha\beta}M_{ab}+\eta_{ab}M_{\alpha\beta}) ,\label{eq:1.7}
\ee

We will consider two types of contractions, rescaling in one case the momenta $P_\alpha$ and in the other case the momenta $P_a$. In both cases the boosts will be rescaled. The first contraction is called a Carroll contraction \cite{Levy-Leblond,Bergshoeff:2014jla,Duval:2014uoa,Cardona:2016ytk}
and it is obtained by the following  rescaling, in the limit of $\omega\to\infty$
\bea
&\tilde M_{\alpha\beta}=M_{\alpha\beta},~~~\tilde P_\alpha={\dd \frac 1\omega} P_\alpha,~~~\tilde M_{ab}=M_{ab},~~~\tilde P_a =P_a,&\nn\\
&\tilde B_{\alpha a}= {\dd \frac 1\omega} B_{\alpha a}.&
\eea
Notice that  the commutators of  $M_{\alpha\beta}$ and $M_{ab}$ with the momenta do not change, since they are linear in the momenta. Therefore we will be interested only in the commutators of the new generators $\tilde P_\alpha$, $\tilde P_a$ and $\tilde B_{\alpha a}$. Using 
 eqs.  (\ref{eq:2.5}),  (\ref{eq:2.6}) and (\ref{eq:1.7}) we find
 \be
 [\tilde B_{\alpha a},\tilde B_{\beta b}]=0,~~~[\tilde B_{\alpha a}, \tilde P_\beta] =0,~~~[\tilde B_{\alpha a}, \tilde P_b]=i\eta_{ab}\tilde P_\alpha.\label{eq:1.9}
 \ee
 We will denote the contracted algebra by ${\cal C}_k(1+D)$.
Since we rescale the first $k$ momenta $P_\alpha$ we will call this contraction a "$k$-contraction" of Carroll type. 
The other contraction that will be considered will be obtained by rescaling the $D+1-k$ momenta $P_a$ and this will be called a $D+1-k$-contraction of Galilei type:
\bea
&\tilde M_{\alpha\beta}=M_{\alpha\beta},~~~\tilde P_\alpha= P_\alpha,~~~\tilde M_{ab}=M_{ab},~~~\tilde P_a ={\dd \frac 1\omega}P_a,&\nn\\
&\tilde B_{\alpha a}= {\dd \frac 1\omega} B_{\alpha a}.&
\eea
Again, the interesting commutators are
\be
 [\tilde B_{\alpha a},B_{\beta c}]=0,~~~[\tilde B_{\alpha a}, \tilde P_\beta] =-i\eta_{\alpha\beta}P_a,~~~[\tilde B_{\alpha a}, \tilde P_a]=0.\label{eq:1.11}
 \ee
This scaling is suggested by the non-relativistic limit of
relativistic branes \cite{Gomis:2000bd,Danielsson:2000gi,Garcia:2002fa,
Brugues:2004an,Gomis:2004pw,Gomis:2005pg}.
The algebra obtained by this contraction will be denoted by ${\cal G}_{D+1-k}(1+D)$. The two algebras ${\cal C}_k(1+D)$ and  ${\cal G}_{D+1-k}(1+D)$ are dual one to the other, they go one into the other by exchanging the role of the momenta $P_\alpha$ and $P_a$ 
and of the metric tensors.

It should be noticed that  this duality  relating   the Carroll and the Galilei  contractions corresponds to quite different physical situations  (see also in the following). In the case of the 1-contraction a similar
duality has been discussed in
 \cite{Duval:2014uoa} although it is not quite the same.

\section{$k$-contractions in configuration space}\label{sec:3}

In this Section we will study the realization of the previous abstract algebras in a
 flat Minkowski space-time, $M(1,D)$, of dimensions $D+1$, with coordinates and metric
\be
x^\mu,~~~\mu=0,1,\cdots,D, ~~~\eta_{\mu\nu}=(-,+,\cdots,+).
\ee
Let us consider this space as the direct sum of  two orthogonal subspaces, a Minkowski space-time $M(1,k-1)$, of dimensions $k$, and an euclidean space $E(D+1-k)$ of dimensions $D+1-k$:
\be
 M(1,D)=M(1,k-1)\oplus E(D+1-k). 
 \ee

 The coordinates of the two subspaces will be chosen to agree with what done in the previous Section
  \be
 x^\alpha\in M(1,k-1),~~~\alpha=0,1,\cdots, k-1
 \ee
 and
 \be
 x^a\in E(D+1-k),~~~a=k, k+1,\cdots,D.
 \ee
It is interesting to notice that this division of the space-time is identical to the on used for defining a $k-1$ brane.

Therefore, as we did previously, we will define   two commuting subgroups of the Poincar\'e group in $D+1$ dimensions, $ISO(1,D)$. These are the Poincar\'e group in $k$ dimensions, $ISO(1,k-1)$ and the roto-translations group in $D+1-k$ dimensions ,$ISO(D+1-k)$. Furthermore, there are the generators that mix together the two subgroups, that is the "boosts", $B_{\alpha a}$.

We will consider the Poincar\'e group $ISO(1,D)$ generated by the following vector fields 
\be
M_{\mu\nu}=-i(x_\mu\de_\nu-x_\nu\de_\mu),~~~ P_\mu =-i\de_\mu
\ee
which satisfy the algebra (\ref{eq:1.1}).

According to the split of the space-time illustrated before, the previous generators are split in
 
\be
(M_{\alpha\beta},P_\alpha)\in ISO(1,k-1),~~~(M_{ab},P_a)\in ISO(D+1-k)
\ee
and the "boosts"
\be
B_{\alpha a}=M_{\alpha a}.
\ee
The contractions in configuration space are dual to the ones defined for the momenta (see, for example,  \cite{Cardona:2016ytk} for Carroll and \cite{Brugues:2004an,Brugues:2006yd} for Galilei) as
\bea\label{trans}
&&{\text Carroll-type},~~~\tilde x^\alpha=\omega x^\alpha,~~~\tilde x^a= x^a,\nn\\
&&{\text Galilei-type},~~~\tilde x^\alpha = x^\alpha,~~~\tilde x^a = \omega x^a,\label{eq:2.8}
\eea
in the limit $\omega\to\infty$. 

These two types of contractions are not equivalent.The terminology used here derives from the case of 1-contractions. In that case the two contractions lead respectively to the Carroll and to the Galilei group with vanishing central charge (see, for instance, \cite{Bacry:1968zf}). Therefore, we have a total of $2D$ possible contractions For the case of p-branes,  $p+1$ possible contractions have been considered \cite{Batlle:2016iel}.

\begin{figure}[h]
\centerline{\psfig{file=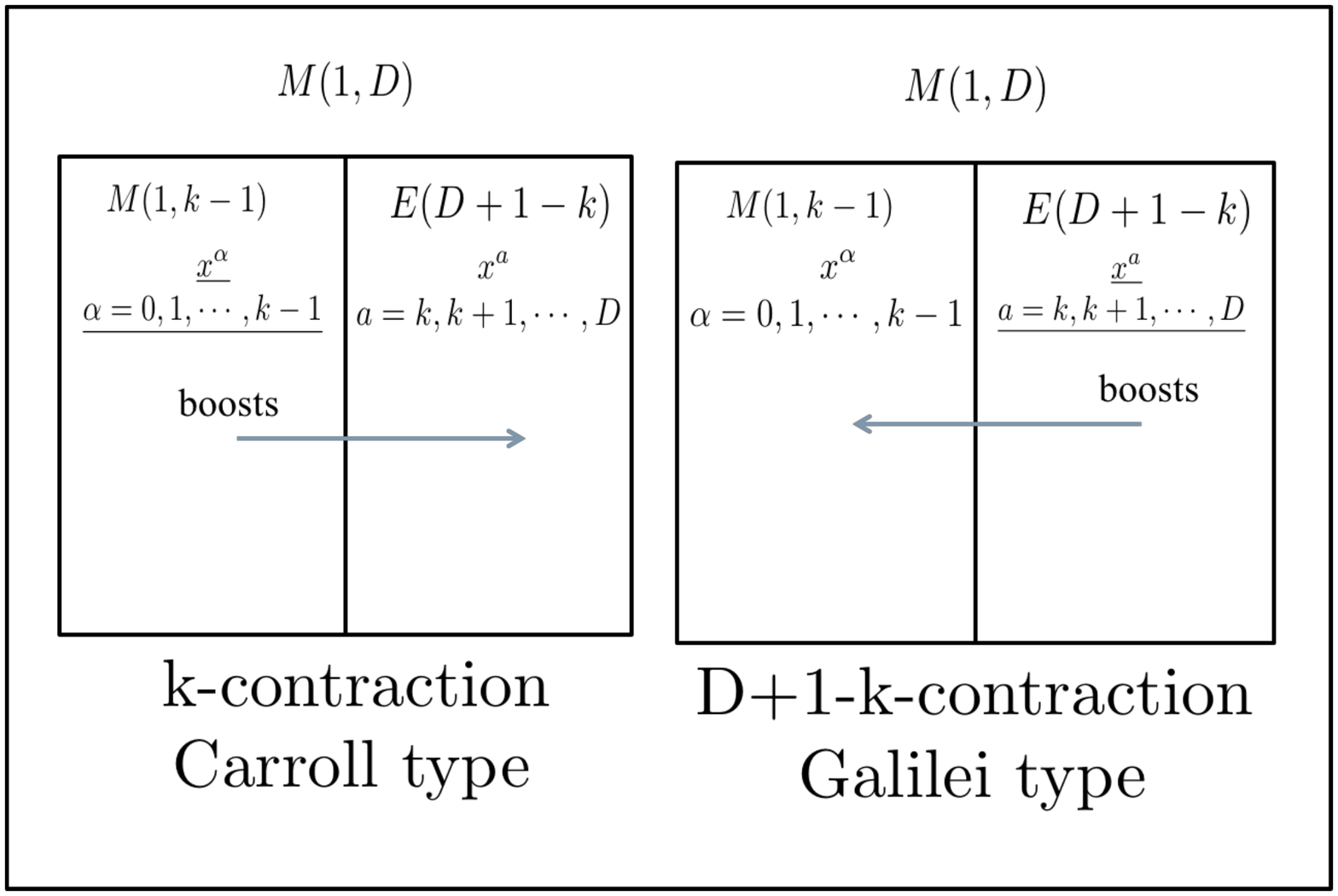,width=10cm}}\noindent
\caption{The two types of contractions considered in the text. In the left panel the Carroll type. In the right panel the Galilei type. The rescaled variables are underlined. The arrows denote the directions in which the boosts act in the two cases.
}\label{Fig:1}
\end{figure}

On the other hand, the duality property considered in the previous Section at the level of the exchange of the momenta for the two contracted groups, here is expressed in terms of the exchange of the manifolds $M(1,k-1)$ and $E(D+1-k)$ (see Fig. \ref{Fig:1}).

In fact, the case in which all the variables are rescaled ($k=D+1$), we get  the original algebra. This is because the Lorentz generators do not change under rescaling of the coordinates, being homogeneous in the coordinates themselves. As for the momenta, since the commutation relations are linear in the momenta, they are left invariant by any common rescaling of the momenta. From this observation it follows that the contractions 
\bea
&&{\text Carroll-type},~~~\tilde x^\alpha= x^\alpha,~~~\tilde x^a= \frac 1\omega x^a,\nn\\
&&{\text Galilei-type},~~~\tilde x^\alpha = \frac 1\omega x^\alpha,~~~\tilde x^a =  x^a,\label{eq:3.9}
\eea
give the same result as the ones following from (\ref{eq:2.8}).

Let us now consider explicitly the two cases:\\\\
\noindent
{\bf Carroll-type}\\\\

Re-expressing the old variables in terms of the new ones,
(\ref{trans}), we get
\be
M_{\alpha\beta}=-i(\tx_\alpha\tde_\beta-\tx_\beta\tde_\alpha)=\tilde M_{\alpha\beta},~~~
M_{ab} =-i(\tx_a\tde_b-\tx_b\tde_a)=\tilde M_{ab}
 \ee
 and, in the limit $\omega\to\infty$
 \be
 B_{\alpha a}=-i\left(\frac 1 \omega\tx_\alpha\tde_a-\omega\tx_a\tde_\alpha\right)\to\omega \tilde B_{\alpha a},
 \ee
where 
 \be
 \tilde B_{\alpha a}=i\tx_a\tde_\alpha.\label{eq:1.16}
 \ee
 Furthermore
 \be
 P_\alpha =-i\omega\tde_\alpha =\omega \tilde P_\alpha,
 \ee
with
 \be
 \tilde P_\alpha =-i\tde_\alpha,~~~\tilde P_a=-i\tde_a.
 \ee
 We see that this contraction coincides with the one of the previous Section.  Therefore the commutation relations of the vector fields are the same obtained for the abstract generators of 
  ${\cal C}_k(1+D)$.
  
 Let us now  consider the Galilei case:\\\\
\noindent
{\bf  Galilei-type}\\\\
Re-expressing the old variables in terms of the new ones,
(\ref{trans}), we get
\be
M_{\alpha\beta}=-i(\tx_\alpha\tde_\beta-\tx_\beta\tde_\alpha)=\tilde M_{\alpha\beta},~~~
M_{ab} =-i(\tx_a\tde_b-\tx_b\tde_a)=\tilde M_{ab}
 \ee
 and
 \be
 B_{\alpha a}=-i\left( \omega\tx_\alpha\tde_a-\frac 1\omega\tx_a\tde_\alpha\right)\to\omega \tilde B_{\alpha a},
 \ee
 in the limit $\omega\to\infty$, with
 \be
 \tilde B_{\alpha a}=-i\tx_\alpha\tde_a.\label{eq:3.17}
 \ee
 Furthermore
 \be
 P_a =-i\omega\tde_a = \omega\tilde P_a,
 \ee
with
 \be
 \tilde P_a =-i\tde_a,~~~\tilde P_\alpha=-i\tde_\alpha.
 \ee
 For the commutation relations, the  considerations made in the Carroll case can be repeated here.
 In any case, for completeness we repeat here the relevant commutators for the two cases:\\\\
\noindent
{\bf Carroll-type}
\be
\left(\begin{array}{c|ccc}
 & B_{\beta b} & P_\beta & P_b \\
\hline
B_{\alpha a} & 0 & 0 & i\eta_{ab}P_\alpha \\
P_\alpha & 0 & 0 & 0 \\
P_a & -i\eta_{ab}P_\beta & 0 & 0
\end{array}\right),
 \ee
 \\\\
\noindent
{\bf Galilei-type}
\be
\left(\begin{array}{c|ccc}
 & B_{\beta b} & P_\beta & P_b \\
\hline
B_{\alpha a} & 0 & -i\eta_{\alpha\beta}P_a &  \\
P_\alpha & i\eta_{\alpha\beta}P_a & 0 & 0 \\
P_a & 0 & 0 & 0
\end{array}\right).
 \ee

\section {Building Dynamical Models}\label{sec:4}

We will consider here the case of  dynamical systems described by discrete variables and then by continuous variables.  The first case corresponds to having a certain number of interacting point-like objects. The second case corresponds to having extended objects as, for instance, branes.
We will show how to construct models invariant either under Carroll, or under Galilei.
 However, we will not consider here the 
  case of field theories.

\subsection{Discrete models}\label{sec:4.1}
Let us start from the  Carroll type of symmetry and consider  a $k$-contraction.
We  suppose to have an action, describing $N$ interacting particles,  invariant under a 
 a linear realization
 Poincar\'e group in $k$ dimensions:
\be
S_{PL}=\int d\tau\, L_{PL}(\dot x^\alpha_i , x^\alpha_i),~~~x^\alpha_i\in M(1,k-1),
\ee
where the index $i=1,\cdots,N$ describes the type of particle. From the invariance under translations, $x^\alpha_i\to x^\alpha_i +a^\alpha$  it follows
\be
\sum_i q_{i\alpha}=0,~~~ q_{i\alpha}=\frac{\de L_{PL}}{\de x^\alpha_i}.\label{eq:4.3}
\ee

 \begin{figure}[h]
\centerline{\psfig{file=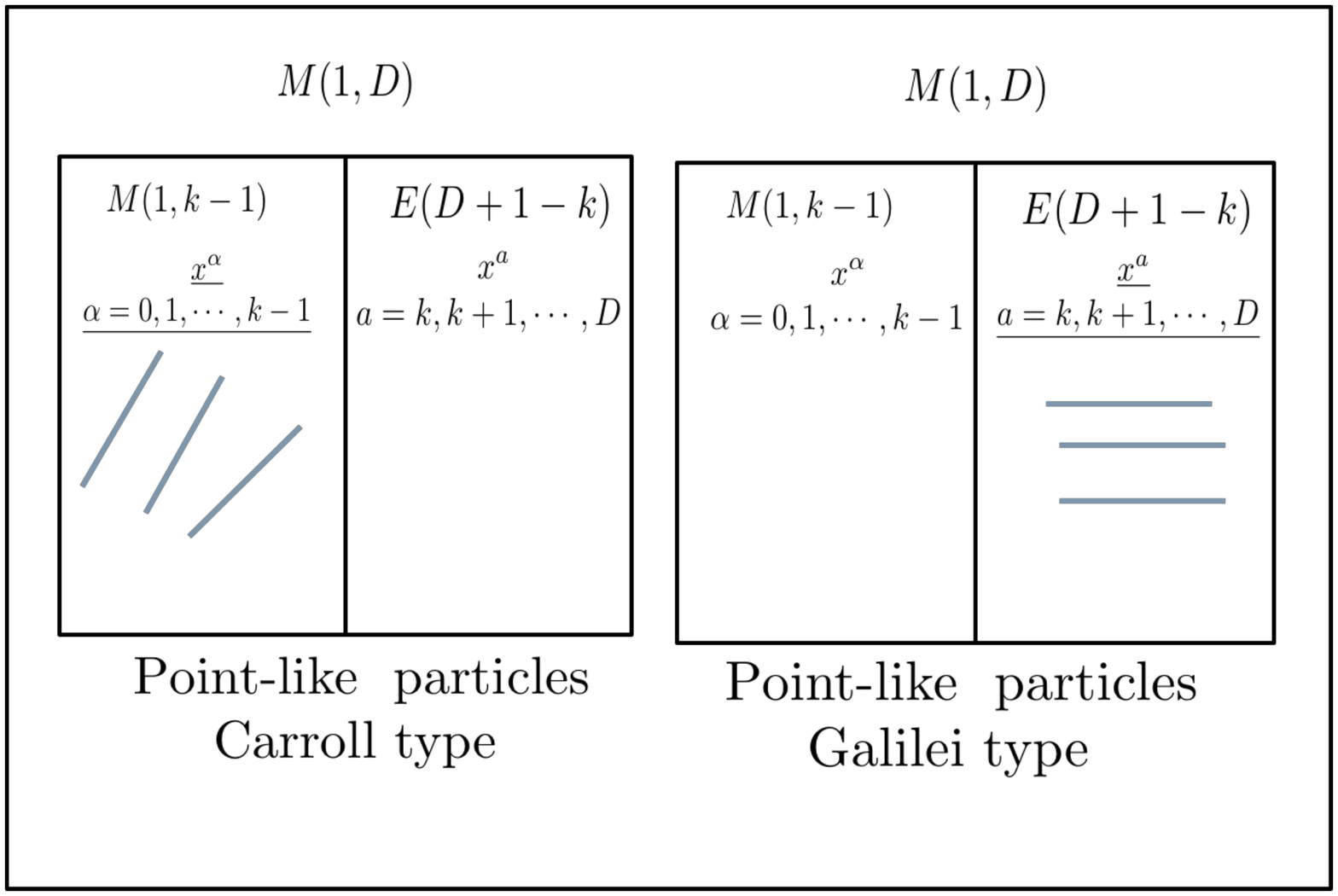,width=10cm}}\noindent
\caption{The two types of point-like systems considered in the text. In the left panel the Carroll type. In the right panel the Galilei type. The lines are a sketch of the world-lines of  particles living  in the  $M(1,k-1)$ Minkowski subspace (left panel) and of the world-lines of the  instantons living in the  euclidean subspace $E(D+1-k)$ (right panel) .} 
\end{figure}

Introducing the canonical momenta
\be
p_{i\alpha} =\frac{\de L_{PL}}{\de \dot x^\alpha_i},
\ee
the equations of motion are
\be
\dot p_{i\alpha}= q_{i\alpha},\label{eq:4.5}
\ee
implying the conservation of the total momentum
\be
\dot P_\alpha= \sum_{i=1}^N \dot p_{i\alpha} = \sum_{i=1}^N q_{i\alpha}=0.\label{eq:4.6}
\ee

Now, let us consider the other part of the total space-time $M(1,D)$, that is the euclidean space $E(D+1-k)$. We will consider our point particles living only in the space $M(1,k-1)$ (see Fig. 2), therefore  we will introduce a set of constraints,  confining the particles to stay in $M(1,k-1)$. Furthermore, if we want to implement the Carroll type of symmetry we have to require that this confinement action is invariant under $ISO(D+1-k)$.
 In particular, we have to require invariance under translations. 
This can be realized  introducing the center of mass coordinates in the euclidean sector 
\be
\bar x^a=\frac 1 N\sum_{i=1}^N x_i^a. 
\ee
Notice that any linear combination $\sum_i^Na_i x_i^a$, with $\sum_i^N a_i=1$, could  also make the job.
 In order to get the momentum conservation the confining action must depend only on the relative coordinates
\be
S_{\rm confining}=\int d\tau\sum_{i=1}^N\left(\lambda^{i}_a \dot x^a_i +\mu^i_a (x^a_i-\bar x^a)\right).
\label{eq:4.7}
\ee

 Notice that we have the identity
 \be
 \sum_{i=1}^N\mu^i_a (x^a_i-\bar x^a) =  \sum_{i=1}^N (\mu^i_a-\bar\mu_a)x^a_i,
 \ee
 where
  \be
  \bar \mu_a=\frac 1 N\sum_{i=1}^N \mu_a^i.\label{eq:4.8a} \ee
 The equations of motion for the variables $x^a_i$ are 
 \be
 \dot p_{ia}=\frac d{d\tau}\frac{\de L}{\de \dot x^a_i}=\frac{\de L}{\de x^a_i}=(\mu_a^i-\bar\mu_a),
 \ee
 implying the conservation of the total momentum $P_a$:
 \be
 \dot P_a=\sum_{i=1}^N \dot p_{ia}=0.
 \ee

 In this way we enforce the translation invariance in $E(D+1-k)$.
 Furthermore, assuming that all the lagrange multipliers transform like vectors under the rotation group $SO(D+1-k)$, the confining action satisfies all our requirements. The equations of motion resulting from this action are
\be
\dot x^a_i =0,~~~x_a^i=\bar x_a. \label{eq:4.9}
\ee
We will now assume as total action
\be
S=S_{PL}+S_{\rm confining}.
\ee
Since the two actions $S_{PL}$ and the confining one have separated variables, all the relations we have derived so far, continue to hold for the total action. Notice that the role of $S_{\rm confining}$ is just to constrain the particles living in $M(1,k-1)$ to not  escape from this space, they are "confined". We will show now that the total action has more symmetries than $ISO(1,k-1)\otimes ISO(D+1-k)$. In fact, it is invariant under the Carroll algebra ${\cal C}_k(1+D)$. 
We have only to check the invariance under the boosts which are given by (compare with eq. (\ref{eq:1.16}))
\be
B_{\alpha}^a=-\sum_{i=1}^N x_{ i}^a p_{i\alpha}.
\ee
A boost with parameters $v^{\alpha}_a$ will generate the transformation
\be
\delta x^\alpha_i=\{\sum_{b\beta}v^{\beta }_b B_{\beta }^b, x^{\alpha }_i 
\}= v^{\alpha }_a x^a_i
\ee
and 
\be
\delta x^a_i=0,
\ee
where we have used the Poisson brackets
\be
\{x^\alpha_i, p_{\beta j}\}=\delta^\alpha_\beta \delta_{ij}.
\ee
Let us now evaluate the variation of the action $S$ under a Carroll boost. We have
\be 
\delta S={\dd\int d\tau\sum_{i=1}^N\left(p_{i\alpha}  v^{\alpha }_a \dot x^a_i+q_{i \alpha}v^{\alpha }_a x^a_i +\delta\lambda^{i}_a \dot x^a_i +\delta\mu^i_a (x^a_i-\bar x^a)
\right)}.\ee
This variation vanishes if 
\be
\delta\lambda^{i}_a=-p_{i\alpha}  v^{\alpha }_a,~~~\delta\mu^i_a=-q_{i \alpha}v^{\alpha }_a.
\ee
Notice that from eq. (\ref{eq:4.6}), due to the translational invariance of $S_{PL}$, it follows
\be
\sum_{i=1}^N \delta\mu_a^i=-\sum_{i=1}^N q_{i \alpha}v^{\alpha }_a=0,
\ee
implying that also the variation of the term proportional to $\bar x^a$ is zero.

Therefore the variations of the lagrange multipliers $\lambda^i_a$ and $\mu^i_a$  are consistent with the translational invariance of the confining action, as it follows from (\ref{eq:4.7}).
 This is not accidental:  the boost invariance   of   $S$ and the translation invariance of  $S_{PL}$      and 
 $S_{\rm confining}$ are strictly related.  In fact, since from Noether's theorem follows that continuous symmetries imply constants of motion, we see that if boosts and translational invariance  in $E(D+1-k)$ are satisfied, then $S_{PL}$ must be translational invariant, as it follows from the commutation relations
 \be
 [B_{\alpha a},P_b] =i\eta_{ab} P_\alpha,
 \ee 
 and the fact that the commutator of two constants of motion is a constant of motion.
 
 We can check  directly that the boosts are conserved quantities using  the equations of motion   (\ref{eq:4.9}), (\ref{eq:4.5}) and (\ref{eq:4.3})
 \be
 \frac d{d\tau} B_{\alpha a} = -\sum_{i=1}^N \dot x_{ i}^a p_{i\alpha}-\sum_{i=1}^N  x_{ i}^a \dot p_{i\alpha}=
 -\sum_{i=1}^N  \bar x^a q_{i\alpha}=0.
\ee
Our   $S_{\rm confining}$ action leads to the following  primary constraints
\be
\phi_{ia}=p_{ia}-\lambda^{ i}_a=0,~~~~\pi_{\lambda^{ i}_a}=0\label{eq:4.21}
\ee
and $\pi_{\mu_a^i}=0\nn.$,
since the action does not contain the time derivatives of the lagrange multipliers.  Of course, other constraints could arise from $S_{PL}$
The first constraint is consistent with the variation of the lagrange multipliers under a boost. In fact
\be
\delta p_{ia}=-\{\sum_{j=1}^N v_b^\beta x_j^b p_{j\beta}, p_{ia}\}=-v_a^\alpha p_{i\alpha}=\delta\lambda^{i}_a.
\ee
The constraints (\ref{eq:4.21}) are second class, in fact
\be 
\{\phi_{ia},\pi_{\lambda^{ j}_b}\}=-\delta_{ij}\delta_a^b.
\ee
Therefore, introducing Dirac brackets, $\{.,.\}^*$, the lagrange multipliers $\lambda^{i}_a$ and the momenta $p_{ia}$ can be identified, since for any dynamical variable, $A$, we have
\be
\{\lambda^{i}_a,A\}^*=\{ p_{ia},A\}^*.
\ee
To evaluate the canonical lagrangian in the reduced phase space, where we have eliminated  $\lambda^{ i}_a$, we first notice that the terms $\lambda^{i}_a \dot x^a_i$ do not contribute to the hamiltonian, being homogeneous of first degree in the time derivative. Therefore
\be
L=\sum_{i=1}^N \left(p_{i\alpha}\dot x^\alpha_i+p_{i a}\dot x^a_i+\mu^i_a (x^a_i-\bar x^a)\right) -H_{PL},
\ee
where $H_{PL}$ is the hamiltonian evaluated from $L_{PL}$. 
 If the lagrangian implies some constraints as in the case of gauge invariance,
for example invariance under  diffeomorphisms, one needs to use the Dirac hamiltonian.
This action is boost invariant under the transformations
\be
\delta x^\alpha_i= v^{\alpha }_a x^a_i,~~~\delta x_i^a=0,~~~\delta p_{ia}=-v_a^\alpha p_{i\alpha},~~~ \delta p_{i\alpha}=0,
\ee
with
\be
\delta \mu^i_a = v_a^\alpha\frac{\de H_{PL}}{\de x^\alpha_i}.
\ee
The Galilei case, dual to the previous one, can be discussed exactly along the same lines. This time the action for the point-like particles is defined in the euclidean space dual to the Minkowski space of the Carroll case:
\be
S_{PL}=\int d\tau L(\dot x^a_i , x^a_i),~~~x^a_i\in E(D+1-k)
\ee
and the confining term defined in $M(1,k-1)$
\be
S_{\rm confining}=\int d\tau\sum_{i=1}^N\left(\lambda^{i}_\alpha \dot x^\alpha_i +\mu^i_\alpha (x^\alpha_i-\bar x^\alpha)\right),
\ee
with
\be
\bar x^\alpha=\frac 1 N\sum_{i=1}^N x_i^\alpha.
\ee
Under boost we have the transformations
\be
\delta x^a_i= v^{a }_\alpha x^{\alpha}_i,~~~\delta x_i^{\alpha}=0,~~~\delta p_{i\alpha}=-v_{\alpha}^a p_{i\alpha},~~~ \delta p_{ia}=0, ~~~ \delta \mu^i_{\alpha} = 
v_{\alpha}^a\frac{\de H_{PL}}{\de x^a_i}=0.
\ee

We will call the Carroll  and the Galilei cases dual one to the other.
 A simple way to get this result, is to start with an action Poincar\'e invariant in the total space $M(1,D)$, say
$S_T$ and define the actions in the two subspaces as
\be
(S_{PL})_{\rm Carroll}=(S_T)|_{x^a\equiv 0},~~~(S_{PL})_{\rm Galilei}=(S_T)|_{x^\alpha\equiv 0}.
\ee
In other words, the point-like actions for the two cases are obtained by restricting the action $S_T$ to the two respective subspaces.

Of course, this is is a quite general way of proceeding in order to get dual models. That is starting from an invariant action in the total space and restricting it to the two subspaces.

\subsection{Continuous models}\label{sec:4.2}
 
We would like to consider the dynamics of extended objects embedded in a confined region of the space-time. We will show that the space-time symmetries of these models are precisely the ones deriving from the $k$-contractions. 
Let us begin considering the Carroll-type of contractions  \cite{Levy-Leblond,Bergshoeff:2014jla,Duval:2014uoa,Cardona:2016ytk}.

An extended object (for instance a brane) is mathematically described by mappings of a manifold (world-sheet) to a target space. Let us start considering the target space as a flat space-time in $D+1$ dimensions. Then, suppose to confine the extended object in the $M(1, k-1)$ Minkowski subspace of the space-time. Assume that the world-sheet is described by the coordinates $(\tau,\sigma_i)$, with $i=1,2,\cdots,m\le k-1$. We will assume for this extended object  an action invariant under a  a linear representation of  $ISO(1,k-1)$
\be
S_{\rm EO}=\int_V d\tau \prod_{i=1}^m d\sigma_i {\cal L}(\dot x^\alpha,x^\alpha_{, i})\equiv \int_V d\tau L_{\rm EO},
\ee
with $x^\alpha$ the coordinates of the target space $M(1,k-1)$,  $V$  the volume which defines the system  and $x^\alpha_{,i}=\de x^\alpha/\de\sigma_i$. Since we want to confine the extended object inside this space, we will add to this action a term keeping into account this condition
\be
S_{\rm confining}=\int _Vd\tau \prod_{i=1}^m d\sigma_i \left(\lambda_a \dot x^a +\sum_{j=1}^m \mu_a^j x^a_{,j}\right)\equiv \int d\tau  L_{\rm confining},\ee
where $x^a$ are the coordinates of the euclidean target space, and the $\lambda$'s and the $\mu$'s are lagrange multipliers. 
The aim of the confining term is to make vanish all the possible motions or vibrations of the extended object that could end in the space $E(D+1-k)$.

The total action is given by
\be 
S=S_{\rm EO}+ S_{\rm confining}\equiv \int_V d\tau L.
 \ee
 The confining term is invariant under the euclidean group $ISO(D+1-k)$, assuming that the all lagrange multipliers transform as vectors under the rotation group $SO(D+1-k)$.  Let us define the following quantities
 \be
{\cal  P}_\alpha =\frac{\delta  L_{\rm EO}}{\delta \dot x^\alpha},~~~ {\cal Q}_\alpha^i =\frac{\delta  L_{\rm EO}}{\delta  x^\alpha_{,i}}, \ee 
 where ${\cal P}_\alpha$ is the momentum density.
 The equations of motion are
 \be
 \frac{\de{\cal P}_\alpha}{\de\tau}+\de_i {\cal Q}^i_\alpha=0,
 \ee
 from which we have that the total momentum, 
 \be
 P=\int_V\prod_{i=1}^m d\sigma_i {\cal P}_\alpha,
 \ee
 is conserved if , 
 \be
 {\cal Q}^i_\alpha\Big|_{\Sigma=\de V}=0,
 \ee
 on the boundary of the volume $V$. 
 
 Let us now show that this action is invariant under the boosts (see eq. (\ref{eq:1.16})) corresponding to a $k$-contraction of Carroll-type:
 \be
 \delta x^\alpha=  v^\alpha_a x^a,~~~\delta x^a=0.
 \ee
 Let us evaluate the variation of the total lagrangian under the previous transformations (the sum over the repeated indices is understood)
 \bea
 \delta L&=&\int  \prod_{i=1}^m d\sigma_i\left({\cal P}_\alpha\delta \dot x^\alpha+{\cal  Q}_\alpha^i \delta x^\alpha_{,i}+\delta\lambda_a\dot x^a+\delta\mu_a^i x^a_{,i}\right),\nn\\
& =& \int  \prod_{i=1}^m d\sigma_i\left({\cal P}_\alpha  v^\alpha_a \dot x^a + {\cal Q}_\alpha^i v^\alpha_a x^a_{,i}+\delta\lambda_a\dot x^a+\delta\mu_a^i x^a_{,i}\right). \eea 
 Therefore the action is invariant by assuming the following transformation law for the lagrange multiplier
 \be
 \delta\lambda_a =-{\cal P}_\alpha  v^\alpha_a,~~~\delta \mu_a^i=-{\cal Q}_\alpha^i  v^\alpha_a,\label{eq:4.45}
 \ee
and in this way we have shown that this construction leads automatically to models invariant under the Carroll type of contracted groups considered in   Section \ref{sec:2}.  An analogous procedure can be made for the Galilei type.

Other features of these models that can be discussed before specifying $S_{EO}$ We have 
\be 
{\cal P}_a =\frac{\delta  L_{\rm confining}}{\delta \dot x^a}=\lambda_a.\ee

Since in the action the time derivatives of the lagrange multipliers do not appear, it follows that the corresponding momenta vanish:
\be
\pi_{\lambda_a}=\frac{\delta S}{\delta\dot\lambda_a} =0.
\ee
These and the previous ones are second-class constraints. Introducing Dirac brackets, by definition we have  
\be
\{{\cal P}_a,A\}^*=\{\lambda_a,A\}^*,
\ee
for any dynamical variable $A$.
Therefore in the reduced phase space,  the  momenta ${\cal P}_a$ and the lagrange multipliers $\lambda_a$ can be identified.  
Notice that the same boosts generating the transformation  (\ref{eq:4.45}) would generate the following variations for the momenta:
\be
\delta {\cal P}_a=-v_a^\alpha {\cal P}_\alpha,~~~\delta{\cal P}_\alpha =0,
\ee
which is consistent with the variation of $\lambda_a$. The construction of the action in  the reduced phase space goes as in the point-like case.
 
 Analogous models can be considered for the Galilei case. In this circumstance, we would take an action describing an extended object inside the euclidean space $E(D+1-k)$
 \be
S_{\rm EO}=\int_V d\tau \prod_{i=1}^m d\sigma_i {\cal L}(\dot x^a,x^a_{, i})\equiv \int d\tau L_{\rm EO},
\ee
with a confining term
 \begin{figure}[h]
\centerline{\psfig{file=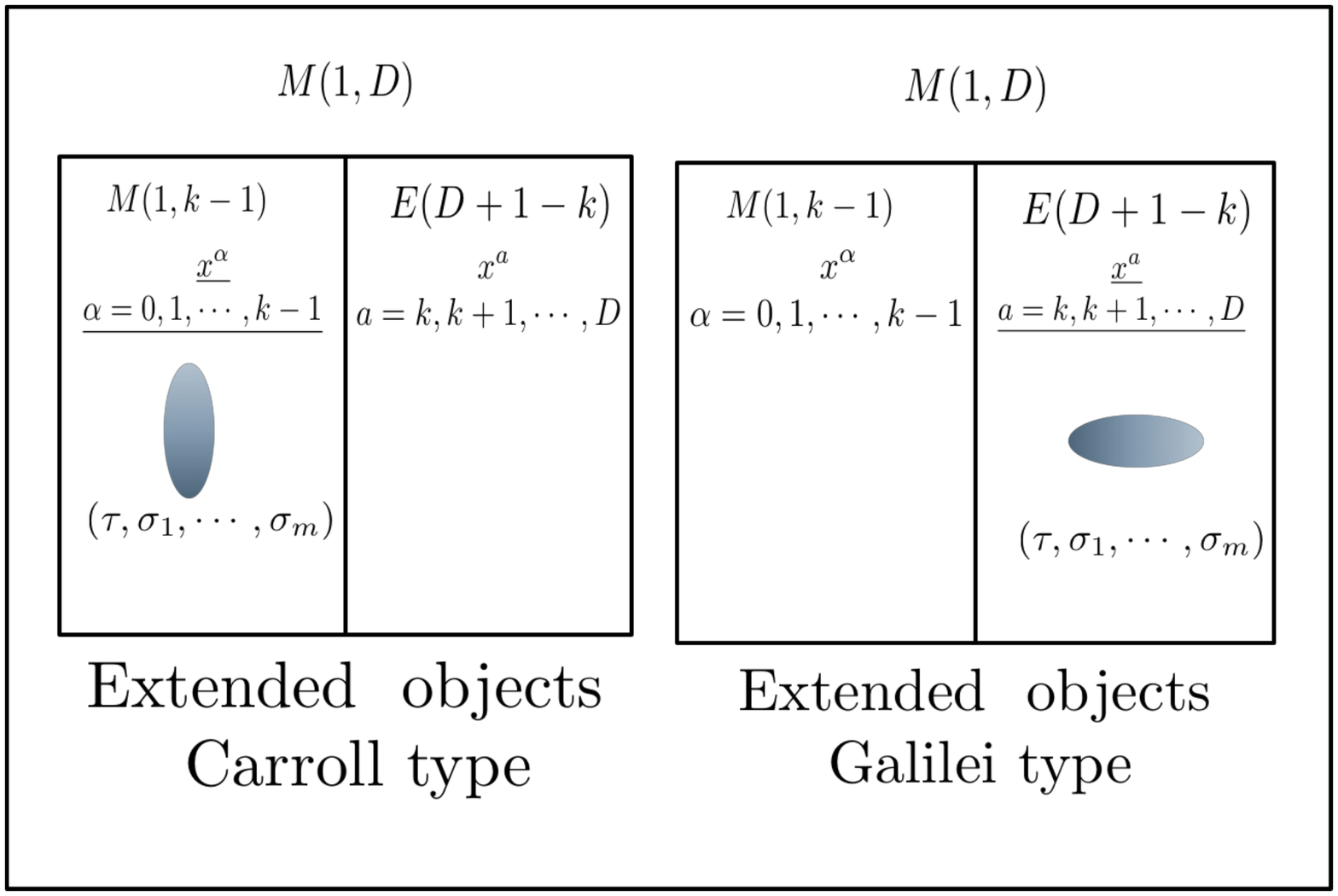,width=10cm}}\noindent
\caption{The shaded figures are a sketch of the extended objects  living  in the  $M(1,k-1)$ Minkowski subspace (left panel) and of the ones  living in the  euclidean subspace $E(D+1-k)$(right panel) .The variables $(\tau,\sigma_1,\cdots,\sigma_m)$ describe the world-sheet of the extended objects. We assume $m\le k-1$ in the Carroll case and $m\le D-k$ in the Galilei case.}\label{Fig:2}
\end{figure}
\vskip0.5cm
\be
S_{\rm confining}=\int_V d\tau \prod_{i=1}^m d\sigma_i \left(\lambda_a^0 \dot x^\alpha +\sum_{j=1}^m \lambda_a^j x^\alpha_{,j}\right)\equiv \int d\tau  L_{\rm confining}.\ee
 Then, one can repeat the same considerations made in the Carroll case. However, there is a very deep physical difference between the two cases, since the Carroll models are  formulated   in a Minkowski space-time, whereas the Galilei case is formulated in an euclidean space. As a consequence the Galilei case resembles the description of an instanton. Notice that  the
 Carroll case corresponding to a $k$-contraction is dual to the $D+1-k$-contraction of Galilei type. However, when we use this duality to describe extended objects there are some conditions to be respected. In fact, we will assume that the dimensions of the world-sheet of the extended object, $m+1$, are smaller or equal to the ones of the target space. Therefore, to have dual objects we have to require $m+1\le k$ , in the Carroll case, and $m+1\le D+1-k$. Therefore to have duality we need the condition (see Fig. \ref{Fig:2})
 \be
 m\le {\rm min} (k-1,D-k).\label{eq:4.49}
 \ee 
 Also in the continuous case we can proceed as in the point-like one, that is starting from an invariant action describing an extended object in $M(1,D)$ Minkowski space. Then, the action for  Carroll  is obtained by restricting this action to the space $M(1,k-1)$, that is by putting to zero the variables $x^a$. Analogously, in  the Galilei case, one takes the restriction to $E(D+k-1)$, that is by putting to zero the variables $x^\alpha$. Of course, in order to get dual models, the condition (\ref{eq:4.49}) must be satisfied.

 \section{Examples}\label{sec:5}
 
 In this Section we will start from the description of  a massive relativistic particle in the full space-time $M(1,D)$ described by the invariant Diff action

 \be
 S=-M\int d\tau\sqrt{-\eta_{\mu\nu} \dot x^\mu\dot x^\nu}.\label{eq:5.1}
 \ee
We will consider various examples. In particular we will study, in  the  case of 1-contraction of Carroll type,
 a massive particle and its Galilei dual, obtaining the Carroll particle studied in \cite{Bergshoeff:2014jla}. Its dual corresponds to the so called Galileian massless particle \cite{Souriau,Batlle:2017cfa}. Notice that, here, massless refers to the fact that the corresponding representation of the  Galilei algebra is with zero central charge. We will consider also a relativistic massless particle described by the action
  \be
  S=\int d\tau \frac 1{2e} \,\dot x^\mu\dot x^\nu\eta_{\mu\nu}. \label{eq:5.2} 
  \ee

 \subsection{The Carroll massive particle}
 
 We will consider the Carroll type 1-contraction \cite{Levy-Leblond}, using the action (\ref{eq:5.1}). The Minkowski space-time reduces to a one dimensional space described by the variable $x^0$. 
 The total action is given by the restriction of  (\ref{eq:5.1}) at this space, plus the confining term 
\be
S=\int d\tau\left(-M\sqrt{(\dot x^0)^{\,2}}\, +\vec\lambda\cdot \dot {\vec x}\right)\equiv \int d\tau L.\label{eq:1.5}
\ee
This  action  is invariant under diffeomorphisms in $\tau$. The momenta are given by
\be
\vec p=\frac{\de L}{\de\dot{\vec x}}=\vec\lambda,~~~p_0=-M\frac {\dot x^0}{\sqrt{(\dot x^0)^{\,2}}}=-M\epsilon(\dot x^0).\label{eq:1.6}
\ee
There is one first-class constraint 
\be
\phi=p_0^2-M^2=0,
\ee
and $2D$ second-class constraints 
\be
\phi_i= p_i-\lambda_i=0,~~~\psi_i=\pi_{\lambda_i}=0,
\ee
with  Poisson brackets 
\be
\{\phi_i,\psi_j\}=-\delta_{ij}.
\ee

Introducing Dirac brackets, one can eliminate the lagrange multipliers $\vec\lambda$ in favour of the momenta $\vec p$. The action (\ref{eq:1.5}) is invariant under the Carroll boost  transformations (see  eq. (\ref{eq:1.16})) with  parameters $\vec v$
\be
\delta x^0= \vec v\cdot\vec x,~~~\delta\vec\lambda =- \vec v p_0,~~~ \delta\vec x=0.\ee

In the reduced phase space, these transformations become 
\be
\delta x^0= \vec v\cdot\vec x,~~~, \delta\vec x =0,~~~\delta p_0 =0,~~~\delta \vec p = -\vec v\, p_0.\label{eq:5.9}
\ee

These transformations are generated by
\be
B_{0a}=-x_a p_0,
\ee
corresponding to the vector fields of eq. (\ref{eq:1.16}). The action in the reduced phase space is
\be
S=\int d\tau\left(\vec p\cdot\dot{\vec x} +p_0\dot x^0 -\frac e 2(p_0^2-M^2)\right).
\ee

This action  is invariant under the boost transformations given in (\ref{eq:5.9}) and coincides with the phase space action studied in \cite{Bergshoeff:2014jla}, where it has been obtained from the relativistic action (\ref{eq:5.1}) in the limit of zero light-velocity. From our point of view, the limit $c\to 0$ is equivalent to the Carroll 1-contraction of eq. (\ref{eq:2.8}). Notice that the physical description of this particle corresponds to a massive particle in its rest-frame. In fact, from the variation of $\vec\lambda$ , we get $\dot{\vec x}=0$.

 \subsection{The Galilei particle}

 The $D$-contracted 
 Galilei case corresponds to the model described in the literature as the Galileian massless particle 
 \cite{Souriau,Duval:2009vt,Batlle:2017cfa}.  The action  is obtained by the restriction of the action (\ref{eq:5.1}) to the euclidean space $E(D)$ and adding the confining part.  Notice that in order to get a real action we have changed the sign inside the square root and, for convenience, also the sign in front of it. Therefore the action that we choose is
\be
S=\int d\tau\left(M\sqrt{\dot x^a\dot x^b \eta_{ab}}+\lambda \dot x^0\right)=\int d\tau\left(M\sqrt{\dot{\vec x}^{\,2}}+\lambda \dot x^0\right)\equiv \int d\tau L,~~~ \eta_{ab}=\delta_{ab}.
\ee
This action is invariant under diffeomorphisms in $\tau$, therefore the canonical hamiltonian vanishes identically. The momenta are given by
\be
\vec p=\frac{\de L}{\de\dot{\vec x}}=M\frac{\dot{\vec x}}{|\dot{\vec x}|},~~~p_0=\frac{\de L} {\de\dot x^0}=\lambda..\label{eq:0.6}
\ee

Therefore there are three  constraints
\be
\phi=\vec p^{\,2} -M^2=0,~~~\phi_1=p_0-\lambda=0,~~\phi_2=\pi_\lambda =0,
\ee
since the time derivative of $\lambda$ does not appear in the action. We see that the first constraint is first-class, whereas the other two are second-class. In fact, their Poisson bracket is not zero
\be
\{\phi_1,\phi_2\}=-1.
\ee
Therefore, introducing the Dirac brackets  in the reduced phase space,  $p_0$ and $\lambda$ can be identified.
  This 
   action  is invariant under the boost   transformations of Galilei type generated by the Galilei boosts (see eq. (\ref{eq:3.17})), with parameters $v_a$.
\be
B_{a0}= x_0 p_a,
\ee

\be
\delta \vec x = \vec v  x^0,~~~\delta\lambda = -\vec v\cdot\vec p.
\ee
In the reduced phase space, these transformations become
\be
\delta \vec x = \vec v  x^0,~~~\delta\lambda =\delta p_0= -\vec v\cdot\vec p,
\ee
and the action is is given by
\be
S=\int d\tau\left(\vec p\cdot\dot{\vec x} +p_0\dot x^0 -\frac e 2(\vec p^{\,2}-M^2)\right).
\ee

  This action coincides with the one studied in   \cite{Souriau,Duval:2009vt,Batlle:2017cfa}. The particle described by this model can be seen as a tachyon in the standard frame of its velocity, that is the frame where $\dot x^0=0$, as it follows varying $\lambda$ in the action. The model  can also be obtained by the contraction  in eq.    (\ref{eq:3.9}), in the limit $\omega\to\infty$  for a relativistic tachyon.

   This is equivalent to consider the limit $c \to\infty$. Therefore, this particle exists at a single instant of time at any point of space. Since this model can be also obtained via  Wick rotation  on the action of a massive relativistic particle, technically it describes an {\bf instanton}. Notice also that the condition $\dot x^0=0$ implies that the physical velocity  $d\vec x/d x^0\to \infty$. 
  
%
%

 \subsection{A  light-like particle of Galilei  type}
 Let us consider the Poincar\'e group in $D+1$ dimensions. If we introduce light cone variables, we can write the generators as 
 
 \be
 P_-~~~, P_+~~~, ~P_a,~~~, M_{-+}=M_{01}, ~~~M_{ab}, ~~~B_{+ a}, 
 ~~~B_{- a},~~~a, b=2,3,\cdots,D,
\ee
where
\be
P_\pm=\frac 12(P^0\pm P^1),~~~ B_a^\pm=(B_{0a}\pm B_{1a}).
\ee
 We can see that we have two subalgebras
 \be{\cal G}^\pm : (M_{01},P_\pm, M_{ab}, P_a, B_a^\pm).
\ee
 These subalgebras have the property that leave invariant a null direction $n^\mu$,
 explicitly $\Lambda^\mu{}_{\nu}n^\nu=\lambda  n^\mu$.
$ {\cal G}^+$ leaves  invariant the direction
 $n^\mu_+=\delta^\mu_+$,whereas $ {\cal G}^-$ leaves invariant the direction   
 $n^\mu_-=\delta^\mu_-$.  
 In the case of  $D+1 =4$,  this algebras are known 
 as $ISim(2)^{\pm}$. They are  symmetries of  the Very Special Relativity \cite{Cohen:2006ky} in place of the Poincar\'e group.
 
The relevant non vanishing commutation relations are
\be
[B_a^\pm, P_\pm]=-iP_a,~~~ [B_a^\pm, M_{01}]=\pm i B_a^\pm, ~~~[M_{01}, P_\pm]= \pm i P_\pm.\label{eq:5.25}
\ee

 In order to construct a a massless particle model we consider a  Galilean invariant model, it is useful to consider a $k=2$ contraction
 of the Poincar\'e group.
 Introducing the light-cone coordinates
\be 
x_\pm=x_0\pm x_1,
\ee
the generators of the contracted algebra in configuration space  (see Section \ref{sec:3}) are 
\bea
&M_{01}= -i (x_+\de_+- x_-\de_-),~~~ P_\pm=-i\de_\pm,~~~ B_a^\pm=-ix_\pm\de_a=x_\pm P_a,\nn\\& \de_\pm=\dd{\frac\de{\de x_\pm}}.&
\eea
 The transformation under the boosts $B_a^\pm$ are
 \be
\delta x^a =-v_\pm^a x_\pm, ~~~\delta x_\pm =0.
\ee
In other words
the main feature of these contracted algebras is that they leave invariant  one of the two branches of the light-cone, $x_\pm=0$, respectively.
Notice that the generator $M_{01}$ acts as a dilation operator on the light-cone variables $x_\pm$. It follows that it leaves invariant one of the planes $x_\pm$ = constant.
This suggests to consider 
 a massless particle in Minkowski space $M(1,D)$ given in (\ref{eq:5.2}) in the light-cone coordinates
Following the lines for a  1-contraction of Galilei type, we will consider two possible actions for describing the particle in the $E(D-1)$ euclidean space
 \be
 S_\pm= \int d\tau \left(\frac 1{2e} \sum_{a=2}^D \dot x_a^2+\lambda_\pm \dot x_\pm\right).
 \ee
These actions are invariant under the two subalgebras
separately. We can check the invariance under the boosts $B_a^\pm$. In fact, we have
\be
\delta S_\pm = \int d\tau\left(-p_a v^a_\pm \dot x_\pm +\delta\lambda_\pm x_\pm\right).
\ee
Therefore $S_\pm$ is invariant assuming
\be
\delta\lambda_\pm =p_a v^a_\pm.
\ee

Let us notice that the two actions $S_\pm$ are invariant under the transformations generated by $M_{01}$. In fact, the corresponding transformations of $x_\pm$ are
\be
\delta x_\pm=\mp s x_\pm,
\ee
where $s$ is the infinitesimal parameter. The invariance follows assuming
\be
\delta \lambda_\pm = \pm s\lambda_\pm.
\ee 
Also this transformation is compatible with the identification of $\lambda_\pm$ with $p^\pm$ as it follows from the commutation relations given in eqs. (\ref{eq:5.25}).

For simplicity let us now consider $S_+$. The same considerations will hold for $S_-$. 
The canonical momenta are, 
 using light cone variables, we have the constraints
\be
p_+=\lambda_+, ~~~, p_-=0, ~~~ {\vec p}^{\,2}=0  \rightarrow \vec {p}=0, ~~~
\pi_{\lambda_+}=0..
\ee
We have $D+1$ first class constraints and two second class constraints, with $2(D+2)$ degrees of freedom.
Then, the  model is described by 2 d.o.f,, $p_+$ and its conjugated coordinate.

The Dirac hamiltonian is given by
\be
H_D=\frac e 2{\,p_a^2}+\mu p_-.
\ee

 \subsection{A  light-like particle of Carroll type}
 
 
{We will consider now a light-like particle within a 2-contraction of Carroll type. The action, in the Minkowski part of the space-time, will depend on the two variables $x^0$ and $x^1$.}


We start with the massless action in eq. (\ref{eq:5.2}) restricted to the plane $(x^0,x^1)$ and we add the confining term:
\be
S=\int d\tau \left[\frac{1}{ 2e}((\dot x^0)^2-(\dot x^1)^2)+\lambda_a\dot x^a\right].\label{eq:2.1}
\ee
The canonical momenta are
\be
p_0=\frac{ \dot x^0}e,~~~p_1=-\frac{ \dot  x^1}e,~~~p_a=\lambda_a,
\ee
from which we have the  first class constraint
\be
\phi=p_0^2-p_1^2=0
\ee
and  $2(D-1)$  second class constraints
\be
\phi_a = p_a-\lambda_a, \chi_a=\pi_{\lambda_a} =0,\,\{\phi_a,\chi_b\}=-\delta_{ab}.
\ee

As in the previous examples, in 
  the reduced phase space (after using the second class constraints), we can identify  $p_a$ with $\lambda_a$.
     We will assume the $\lambda_a$'s transforming like vectors under the generators of $SO(D-1)$). The model  is invariant under translations, and under the transformations generated by $M_{01}$, since this generates a Lorentz boost in the direction $D$ and it leaves invariant the quadratic form in the action. Furthermore, we have invariance under the two type of boosts $B_{0a}$ and $B_{1a}$:
\be
\delta_{B_{0a}} x^0= -v_{0a} x^a,~~~ \delta_{B_{0a}}\lambda_a =p_0v_{0a}
\ee
 and
 \be
 \delta_{B_{1a}}  x^1= -v_{1a} x^a,~~~\delta_{B_{1a}}\lambda_a =p_1v_{1a}. 
 \ee
 The rest of the discussion goes as in the other examples.
In this case the Carroll particle moves in the plane $(x_0,x_1)$ at the speed of light.

In the case of  the 2-contraction we have reported only the case of a massless particle.. We could as well to start with the action for a massive particle in $M(1,D)$. In this case we would obtain for the Galilei particle the mass-shell conditions
\be
p_0^2-p_1^2 =0,~~~\sum_a p_a^2=M^2,
\ee that is a tachyon in $D+1$ dimensions, whereas for the Carroll case
\be
p_0^2-p_1^2 =M^2,
\ee with considerations completely analogous to the ones discussed for the massless case.

\subsection{A model of two particles}

We start we considering the relativistic two- particle model \cite{Dominici:1977fh}
\be\label{oldDGLK} 
L=-\sqrt{-(m_{10}^2-V({r^2}))\,{{\dot x}_1}^2}-
     \sqrt{-(m_{20}^2-V({r^2}))\,{{\dot x}_2}^2}
=-\sum_{j=1,2}  \sqrt{-m_j^2({r^2})\,{{\dot x}_j}^2},
\ee
where ${x_j}(\tau), (j=1,2)$ are the space-time coordinates of the two particles.  
$V({r^2})$ is any  Poincar\'e invariant function of the squared relative distance  
 $r^2=(x_2-x_1)^2$,
$m_{j0}$'s are the rest masses of the particles and 
$m_{j}^2({r^2})=m_{j0}^2-V({r^2})$ are the effective masses of the particles. 
The interaction breaks the individual invariance under diffeomorphism (Diff)
 of the action of two free particles, leaving a universal Diff invariance.
 The momenta are given by
\be
p_i= m_{i}^2({r^2})\frac{\dot x_i}  {\sqrt{-m_i^2({r^2})\,{{\dot x}_i}^2}}.
\ee

Following our procedure the Carroll lagrangian will be given by 
\be\label{oldDGLKnew} 
L=-\sqrt{-(m_{10}^2-V({r_0^2}))\,{(\dot x}_1^0)^2}-
     \sqrt{-(m_{20}^2-V({r_0^2}))\,({\dot x}_2^0)^2}+\vec{\lambda}_1\cdot\dot{\vec x}_1
     +\vec{\lambda}_2\cdot\dot{\vec x}_2+\vec\mu\cdot\vec r,
\ee
where $r^0=x_1^0-x_2^0$.  Like in the $k=1$ Carroll particle, the presence of the second class constraints allows to eliminate the ${\lambda}_j$ in terms of the momenta $p_j$.
In the reduced space the canonical action becomes
\be
S=\int d\tau \left(\sum_{j=1,2}({p_i}_\mu{\dot x_i}^\mu-\frac {e_i}{2}( -p_{0j}^2+m_j^2({r_0^2})))-\vec\mu\cdot\vec r\right).
 \ee
The particles do not move, although the momenta of the particles is not individually conserved.
Notice that for these Carroll particles the momenta are not related to the velocities of the
particles. This model is different from the two particle model of \cite{Bergshoeff:2014jla} where the 
two mass shell constraints depend on the total momenta of the particles.

The Galileian counterpart is given
\be\label{oldDGLKnew} 
L=-\sqrt{(m_{10}^2-V({\vec r^{\,2}}))\,{\dot{\vec x}_1}^2}-
     \sqrt{(m_{20}^2-V({\vec r^{\,2}}))\,{\dot{\vec x}_2}^2}+\lambda_1\dot x^0_1+\mu(x_1^0-x_2^0)
     +\lambda_2\dot x^0_2.
\ee
Like in the $ k=1$ Galilei particle, the presence of the second class constraints allows to eliminate the ${\lambda}_i$ in terms of the momenta $p_{0i}$.
The action canonical in the reduced space is 
\be
S=\int d\tau \left[\sum_{i=1,2}(p_{0i}\dot x_i^0+\vec p_i\cdot\vec x_i-\frac {e_i}{2}( {{\vec p}_i}^{\,2}-m_j^2({\vec r^{\,2}}))-\mu(x_1^0-x_2^0).\right] \ee

The two particles described by this model can be seen as  two tachyons in their standard frame of  their velocity, 

\subsection{2-contraction for a Carroll string}

We recall the string action 
\be
S_{\rm string}=-T\int d\tau\int_0^\pi d\sigma\sqrt{(\dot X\cdot X')^2-(\dot X)^2(X')^2}.
\ee
Following what we have done in Section \ref{sec:4.2} we assume the following action
\be
 S=S_{\rm string}+ \int d\tau \int_0^\pi d\sigma\left(\lambda_a\dot X^a+\mu_a {X'}^{a}\right),
 \ee 
 where in the string action $X^\alpha=(X^0,X^1)$, whereas  $a=2,3,\cdots,D$. We get for the string lagrangian density
 ${\cal L}_{\rm string}$
 \be
{\cal L}_{\rm string}=  -T\sqrt{(\dot X^0X^{1'}-X^{0'}\dot X^1)^2}.
\ee
The component of the  canonical momentum are given by
\be
{\cal P}_0 =-T\frac{(\dot X^0X^{1'}-X^{0'}\dot X^1)}{\sqrt{(\dot X^0X^{1'}-X^{0'}\dot X^1)^2}}
X^{1'},
\ee
\be
{\cal P}_1 =T\frac{(\dot X^0X^{1'}-X^{0'}\dot X^1)}{\sqrt{(\dot X^0X^{1'}-X^{0'}\dot X^1)^2}}
X^{0'}.
 \ee
 We get the primary constraints 
 \be
 {\cal P}_\alpha X^{\alpha '}=0,~~~ {\cal P}^2+  T^2 X^{'2}=0.
 \ee 
 Let us evaluate the quantities ${\cal Q}^\sigma_\alpha={\de{\cal L}}/{\de {X'}^{\alpha }}$
 \be
 {\cal Q}^\sigma_0=T\frac{(\dot X^0X^{1'}-X^{0'}\dot X^1)}{\sqrt{(\dot X^0X^{1'}-X^{0'}\dot X^1)^2}}\dot X^1,
 \ee 
  \be
 {\cal Q}^\sigma_1=-T\frac{(\dot X^0X^{1'}-X^{0'}\dot X^1)}{\sqrt{(\dot X^0X^{1'}-X^{0'}\dot X^1)^2}}\dot X^0.
 \ee
 The canonical action in the reduced space is given by
 \be
S=\int d\tau d\sigma\left ( {\cal P}_{1\mu}\dot X_1^\mu+ {\cal P}_{2\mu}\dot X_2^\mu
-\frac {e}{2}( {\cal P}_\alpha {\cal P}^\alpha+  T^2 X^{'2})
-\mu({\cal P}_\alpha {X'}^{\alpha }) -\mu_a   {X'}^{a}   \right).
 \ee
 This model is different from the Carroll string of \cite{Cardona:2016ytk}.
 
 Choosing the gauge conditions
 \be
 X^0=\tau,~~~~X^1=\sigma,
 \ee
 we have
 \be
 {\cal P}_0=-T ,~~~ {\cal P}_1= 0,~~~ {\cal Q}^\sigma_0=0,~~~ {\cal Q}^\sigma_1= -T,
  \ee
 and for the velocities
 \be
 \dot X^0=1,~~~\dot X^1=0.
 \ee
 These equations  show that the string is at rest and, having fixed end points, it satisfies the Dirichlet boundary conditions.
 The total momentum is conserved
  \be
 P_\alpha= \int_0^\pi d\sigma \,{\cal P}_\alpha, ~~~E= P^0= \pi T,~~~P_1=0.\ee
 Therefore in the case of a 2-contraction of Carroll type, the string is the equivalent of the Carroll particle, that is the case of 1-contraction.  For the considerations about the invariance under the boosts, one can repeat the general considerations made in Section \ref{sec:4.2}

 \section{Conclusions and outlook}
 
 In this paper we have introduced a general method to construct models 
with invariant lagrangians under generalized Carroll or Galilei algebras (in the text called $k$-contractions). The method consists in starting from a space-time in $D+1$ dimensions and partitioning it in two parts, the first Minkowskian and the second Euclidean. Then, a Carroll invariant model can be obtained by introducing a Minkowski invariant action in the first part of the space-time, whereas in the second part a system of lagrange multipliers, transforming in an appropriate way  under the Euclidean group is introduced. This system is such to compensate the variations, induced by the Carroll boosts, of the action previously defined. The same procedure is done for the Galilei case, this time using an    lagrangian defined in the Euclidean sector and enlarging it with a system of lagrange multipliers living in the first part of the space-time. The main difference between Carroll and Galileian models constructed in this way, is that in the first case we have a real dynamical system, since the time coordinates are in the action, whereas in the second case the time variables appear only in the confined part, meaning that one is describing an instanton-like object.
 
 This procedure could be generalized as follows: start with a target space (in the text the original space-time), partitioned  in two parts. Then, suppose that the target space supports a natural representation  of some group $G$ (in the text the realization in terms of vector fields on the space-time). Assume that the two sectors support natural representations of two groups $G_I$ and $G_{II}$, which are both subgroups of $G$. Eventually assume that  the Lie algebra of  $G$ can be decomposed as follows: 
 \be
 {\rm Lie~G}= {\rm Lie~G_I}\oplus {\rm Lie~G_{II}}\oplus I,
  \ee
  where $I$ are a set of intertwining generators mapping ${\rm Lie~G_I}$ into ${\rm Lie~G_{II}}$ or viceversa.
  Then, dynamical models describing systems, living in the target space, can be constructed  following what we have done in the text. All these models would exhibit the confinement in one of the two sectors of the target space.
  
   One could think to various possible extensions. For instance, one could describe a dynamical model with a generalized Galilei invariance, starting with a space-time with two times, $M(2,D-1)$. Or else,  starting from a euclidean space $E(D+1)$ one could consider statistical systems confined to dimensions lower that $D+1$, but exhibiting a bigger symmetry. 
 This approach could also be useful in theories with extra-dimensions.
  We hope to be able to consider in the future some of these extensions in greater detail.
 
 \section*{Acknowledgments}
 
 We acknowledge Eric Bergshoeff, Jaume Gomis and Jorge Zanelli
for comments. JG has been supported in part by FPA2013-46570-C2-1-P and Consolider CPAN and by the Spanish government (MINECO/FEDER) under 
 project MDM-2014-0369 of ICCUB (Unidad de Excelencia Mar\'\i a de Maeztu).
 JG has also been  been supported by  CONICYT under grant PAI801620047 as a
 visiting professor of the Universidad Austral de Chile.  
 
%
%

\end{document}